\documentclass[conference,english]{IEEEtran}  
\usepackage{babel}
\usepackage[utf8]{inputenx}
\usepackage[cmintegrals]{newtxmath} 
\usepackage[caption=false]{subfig}
\usepackage{graphicx}
\usepackage{amsmath} 
\usepackage{amssymb}  
\usepackage{balance}
\usepackage[autostyle]{csquotes}
\usepackage{url}
\usepackage{tabu}
\usepackage{microtype}
\usepackage{booktabs}
\usepackage{textcomp} 
\usepackage{siunitx}
\usepackage{cite} 
\usepackage{listings}
\usepackage{flushend} 
\usepackage{tikz}
\usepackage[hidelinks]{hyperref}  

\newcommand\copyrighttext{%
	  \footnotesize \textcopyright 2017 IEEE. Personal use of this material is permitted.
	    Permission from IEEE must be obtained for all other uses, in any current or future
	      media, including reprinting/republishing this material for advertising or promotional
	        purposes, creating new collective works, for resale or redistribution to servers or
		  lists, or reuse of any copyrighted component of this work in other works.
		    DOI: \href{https://doi.org/10.1109/SDS.2018.8370432}{10.1109/SDS.2018.8370432}
		    }
		    \newcommand\copyrightnotice{%
			    \begin{tikzpicture}[remember picture,overlay]
				    \node[anchor=south,yshift=10pt] at (current page.south) {\fbox{\parbox{\dimexpr\textwidth-\fboxsep-\fboxrule\relax}{\copyrighttext}}};
			    \end{tikzpicture}%
			    }

\DeclareSIUnit{\bit}{b}
\DeclareSIUnit{\byte}{B}

\newcommand{\esp}[1]{\mathrm{E}\left[{#1}\right]}
\newcommand{\Tany}[1]{\mathrm{T}_{\mathrm{#1}}}
\newcommand{\Toff}{\Tany{off}}
\newcommand{\TS}{\Tany{S}}
\newcommand{\TW}{\Tany{W}}
\newcommand{\soff}{\sigma_{\mathrm{off}}}

\title{Implementing Energy Saving Algorithms for Ethernet Link Aggregates with ONOS}

\author{%
  \IEEEauthorblockN{Pablo Fondo-Ferreiro, Miguel Rodríguez-Pérez, Manuel Fernández-Veiga}%
  \IEEEauthorblockA{atlanTTic Research Center\\
    University of Vigo\\36310 Vigo, Spain\\Tel.:+34~986~813459;
    fax:+34~986~812116; email:~\url{pfondo@det.uvigo.es}}%
}

\begin{document}
\maketitle
\copyrightnotice

\begin{abstract}
  During the last few years, there has been plenty of research for reducing energy
  consumption in telecommunication infrastructure. However, many of the proposals
  remain unimplemented due to the lack of flexibility in legacy networks.
  In this paper we demonstrate how the software defined networking (SDN) capabilities of current networking equipment can be used to implement some of these energy saving algorithms.
   In particular, we developed an ONOS application to realize an energy-aware
  traffic scheduler to a bundle link made up of Energy Efficient Ethernet (EEE)
  links between two SDN switches. We show how our application is able to dynamically adapt to the traffic characteristics and save energy by concentrating the traffic on as few ports as possible. 
  This way, unused ports remain in Low Power Idle (LPI) state most of the time, saving energy.
\end{abstract}

\begin{IEEEkeywords}
  SDN, ONOS, IEEE 802.3az, Energy Efficiency
\end{IEEEkeywords}

\section{Introduction}
\label{sec:intro}

\IEEEPARstart{T}{wo} recent trends have been shaking the networking landscape during the
last few years. On the one hand, the seemingly unstoppable deployment of SDN
equipment and solutions in datacenters, has provided network operators with
unprecedented flexibility. On the other hand, growing concerns about both environmental and
monetary costs of ICT infrastructure has led to the proposal of a plethora of
solutions to augment the energy efficiency of networking equipment.

However, a great deal of interesting proposals for reducing energy
usage remain unimplemented because they either require
significant changes to fundamental networking protocols or to the actual
networking equipment, cf.~\cite{Chiaraviglio2012,Jung2014,Kim2012,RodriguezPerezb}.
We believe that software defined networking (SDN) can be used to overcome these
limitations. In particular, having a comprehensive view of the network
status and the ability to precisely control individual flow forwarding, can be 
used to define energy-optimum paths for traversing flows, either
through the whole network domain~\cite{Chiaraviglio2012,Jung2014,Kim2012} or 
when crossing a bundle link between two directly connected switches~\cite{RodriguezPerezb}.

Open Network Operating System (ONOS)~\cite{onos} is an open source project, designed for high availability, performance and scalability. 
ONOS abstracts the particular details of the actual SDN forwarding devices in the network, permitting the development of technology-agnostic network applications. These applications can obtain a global view of the network and they can also modify 
the actual flow tables in real time.

In this paper, we describe an ONOS application for the optimal distribution of network traffic among a bundle of energy efficient Ethernet (EEE)~\cite{ontheroad} links managed by a SDN network so that energy savings are maximized. Our application dynamically identifies the network flows traversing the bundle and adjusts the forwarding tables to achieve the energy optimal share of traffic.

Although there exists an optimal way to share the traffic as a whole among the links~\cite{RodriguezPerezb}, that is the theoretical base of our implementation, the actual assignment of individual flows to the EEE links remains an open problem. Thus, in this paper we have compared three different flow scheduling algorithms. All three algorithms produce a similar share of traffic among the links, so they achieve energy saving results very near those predicted by the model. However, as the particular flows assigned to the links are different, their characteristics regarding traffic delay and loss rate differ.

The rest of this paper is structured as follows:
Section~\ref{relatedwork} presents the related work. Then, we proceed with
the problem statement in Section~\ref{sec:statement}. Section~\ref{sec:sdn-impl}
describes our SDN implementation, while Section~\ref{algorithms} describes some
alternatives for the implementation. Section~\ref{results} summarizes
the results obtained and finally Section~\ref{conclusions} exposes the
conclusions.

\section{Related Work}
\label{relatedwork}

The usage of SDN networks to diminish energy usage in computer networks has 
already been explored by several authors.

A survey on energy efficiency in SDNs is presented in~\cite{tuysuz2017survey}.
This survey analyzes the different components of the SDN structure which can
be dynamically configured to reduce power consumption. The approaches analyzed include reorganizing the flows in the network to have a small number of active devices in the network so that the unused devices can be put into sleep mode. When there is a low traffic
load, they also mention the possibility of putting certain ports rather than whole devices into sleep
mode. Those kind of approaches which set devices in a low-power mode based on
the current load of the system, are usually referred to as traffic-aware.

GreenSDN, a SDN emulation environment based on Mininet and the python based POX
SDN controller, has been proposed in~\cite{rodrigues2015greensdn}, where they
report on the difficulties they faced building a SDN environment with
capabilities of emulating the energy saving protocols operating at different
levels of the network. They propose a mechanism which operates at the node 
level, exploiting the Low Power Idle (LPI) mode defined by IEEE~802.3az
that is especially relevant to this paper. However, they only consider turning 
on and off whole switches and not individual interfaces when the traffic load is behind a bundle.

Another contributions in the literature about energy-efficiency leveraging the power of OpenFlow include ElasticTree~\cite{heller2010elastictree} and ECODANE~\cite{huong2011ecodane} which are data-center based proposals. Both proposals traffic-aware mechanisms consistent in dynamically turning links and devices on and off based on the current traffic load of the system.

The authors in~\cite{RodriguezPerezb} have shown that, under suitable
conditions, the traffic load allocation that minimizes the energy consumption
in a bundle of EEE links is achieved using a water filling algorithm. However,
the proposed algorithm cannot be directly ported to
SDN. First, the proposed water filling algorithm operates at
the packet level, that is, when a switch receives a new packet that must be
sent through the bundle, it has to decide the port used to transmit the packet
based on the current backlog of the port. However, SDN operates at the flow level, therefore, all packets belonging to the same flow will be forwarded via the same port. Secondly, the algorithm needs to obtain the 
queue occupation of each port to classify incoming packets, but unfortunately SDN does not provide access to such information, to the best of our knowledge.

\section{Problem Statement}
\label{sec:statement}

This section describes the algorithm that minimizes the energy consumption in a bundle of EEE links.
We will briefly summarize the results from~\cite{RodriguezPerezb}
describing the optimum traffic allocation in bundled EEE links. The authors there demonstrate that, for certain common class of functions that characterize the energy-consumption profile of the links, the solution to the optimum
allocation for a given offered traffic load is a simple sequential water-filling algorithm: each link capacity is fully used before sending traffic through a new otherwise idle link.

Luckily, the energy-consumption profile of both major modes used to govern the
use of the LPI mode (namely the \emph{frame transmission} and \emph{packet coalescing}
algorithms) belong to this class of functions.
In summary, to achieve the optimum energy savings links will be in the following 
states: some links used to its full capacity, some links completely idle and at 
most one link transmitting packets at less than its full capacity.

Nevertheless, the optimum allocation in terms of maximum energy savings can
easily lead to packet delays growing uncontrolled, if proper care is not
taken. This issue has also been carefully analyzed in~\cite{RodriguezPerezb},
where the authors propose modifications of the simple water-filling algorithm
to control the average delay of the packets with a bounded cost in the energy
savings.

\section{SDN Based Implementation}
\label{sec:sdn-impl}

We now proceed to describe the implementation of the algorithm using the facilities
provided by SDN switches.

The first challenge resides in the fact that while the theoretical solution
described in~\cite{RodriguezPerezb} assumes a packet level operation our 
implementation will have to operate at the flow level.

\subsection{Flow Selection}
\label{flowsdefinition}

Recall that SDN works at the flow level (e.g., OpenFlow~\cite{mckeown2008openflow} switches are composed of flow tables), where a flow is defined by a set of fields of the incoming packets. Therefore, the first step is to define which fields will identify our flows. SDN allows us to define very different levels of flow granularity:
a very coarse level of granularity will only match on the destination MAC
address or the physical input port of the packets, for example. However, this
kind of definition of the flows does not seem to be suitable for our purpose
because if deployed in a transit network, flows will share a common small set of
exiting routers, thus limiting the variability of MAC addresses.

Since such a coarse granularity does not seem to be suitable for our
algorithm, we have to identify a finer level of granularity that may allow us to
split the traffic among the links. Moreover, we strive to aggregate non correlated
flows, so their aggregated behavior is more stable. We have explored several levels of
granularity to identify different flows among the packets so that we can send 
some flows to one port and another flows to other port, and as a result we will
be able to utilize the full capacity of the bundle. We have explored two 
alternatives to achieve this: \emph{flow tagging} and \emph{field matching}.

First, we have considered the possibility of having a process at the
input of our SDN network which smartly tags the packets with a flow
label, imposing the tag assigned to each packet in a field directly matchable
by ONOS (e.g., the DSCP field in the IP header). This way we could have the
packets evenly distributed into a fixed number of flows, which would ease the
allocation of the flows to the bundle. The main drawback of this approach is that
it needs a dedicated tagger at input nodes in the network, so we have also
tried to select flows in a distributed manner.

We would like the aggregated flows to be later assigned to a given port
to show a predictable, ideally constant, demand. To this end, we try to 
aggregate independent end-to-end network layer flows, thus defined by the
source and destination IP addresses pair of each packet. Note that, at layer~2
we could have an insufficient number of identifiable flows (e.g., in a transit network). On the
contrary, the number of transport layer flows can be excessively high.
Consequently, the aggregation of layer~3 flows is expected to result in a
low variance in the rate of the aggregated flows, hence being the rate of
these flows more predictable.

However, a direct mapping between a pair of source and destination IP
addresses to a flow will produce $2^{64}$ different flows, which is clearly an
unacceptable huge number of flows and obviously not scalable. Even considering
only the destination address will produce $2^{32}$ different flows, which is
also unacceptably large. Thus we have chosen use only some bits of the
destination IP address to identify subflows inside the packets destined to the
same MAC address.

\begin{figure}
  \centering \includegraphics[width=\columnwidth]{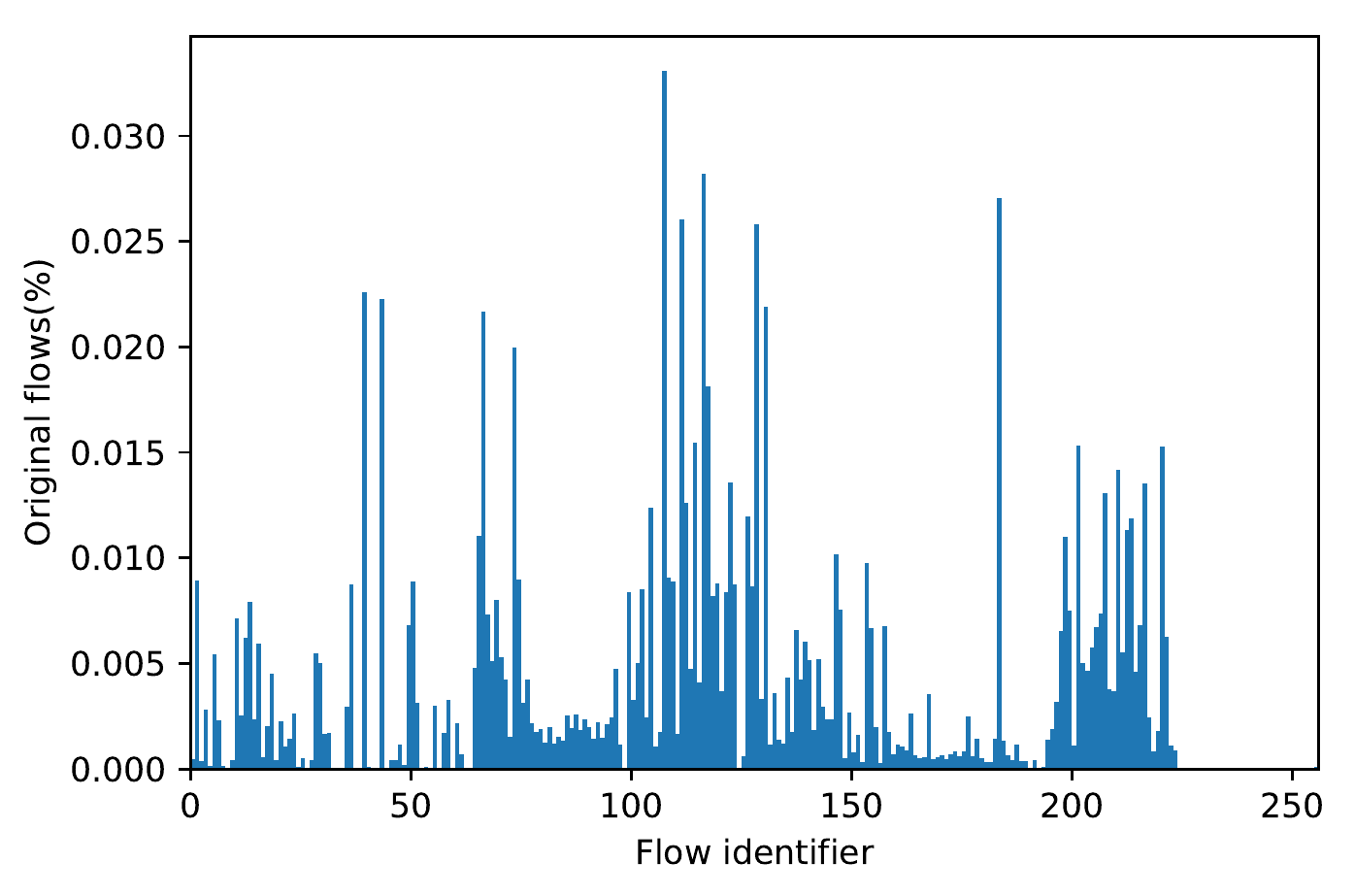}
  \caption{Flow distribution for the 8 first bits of the destination address.}   \label{fig:hist_dst_0_8}
\end{figure}
\begin{figure}
  \centering
  \includegraphics[width=\columnwidth]{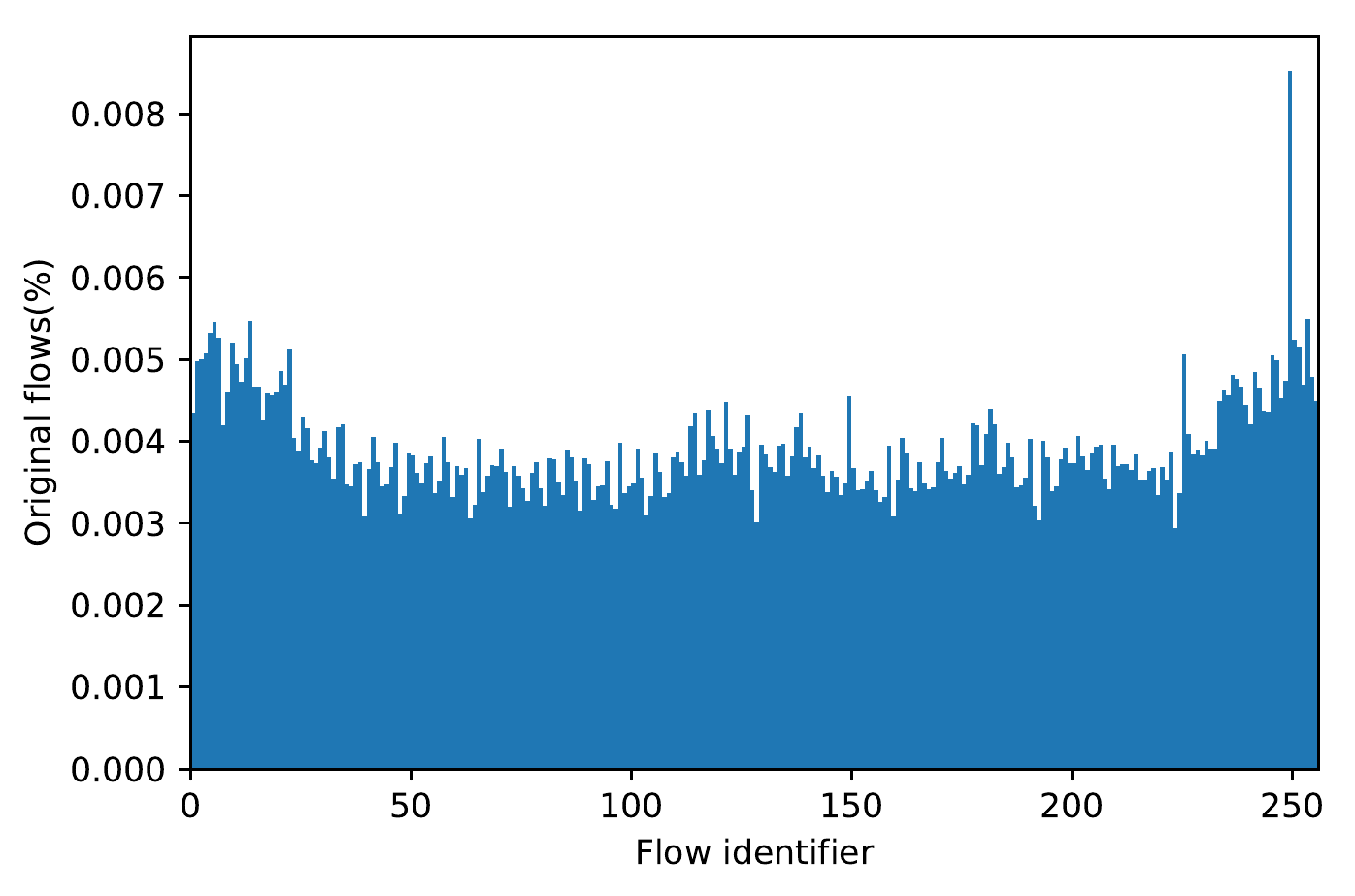}
  \caption{Flow distribution for the 8 last bits of the destination address.}
  \label{fig:hist_dst_24_32}
\end{figure}
Since we want to have a flatter distribution of the original flows (pair of
source and destination IP addresses) to the aggregated flows, we have explored
the usage of arbitrary bitmasks over IP addresses and analyzed how the flows
were distributed.\footnote{The traffic traces analyzed come from the publicly
  available passive monitoring CAIDA dataset~\cite{caida13}.} Specifically, we
have counted how many of the original layer~3 end-to-end flows of our trace
will be contained in each aggregated flow. The flatter results were obtained
for the last bits of the destination IP address, as shown in the histograms of
Figs.~\ref{fig:hist_dst_0_8} and~\ref{fig:hist_dst_24_32} (other combinations
have been tested and show similar results, thus they are omitted for the sake
of brevity).

However, although the current version of OpenFlow allows arbitrarily
bitmasking IP addresses, the ONOS controller does not, and restricts us to
mask the first bits of the IP address. Although using the largest number of bits
yields to lower variance in the distribution of the original end-to-end flows
\begin{figure*}
  \centering 
  \subfloat[Last 4 bits]{\includegraphics[width=.49\textwidth]{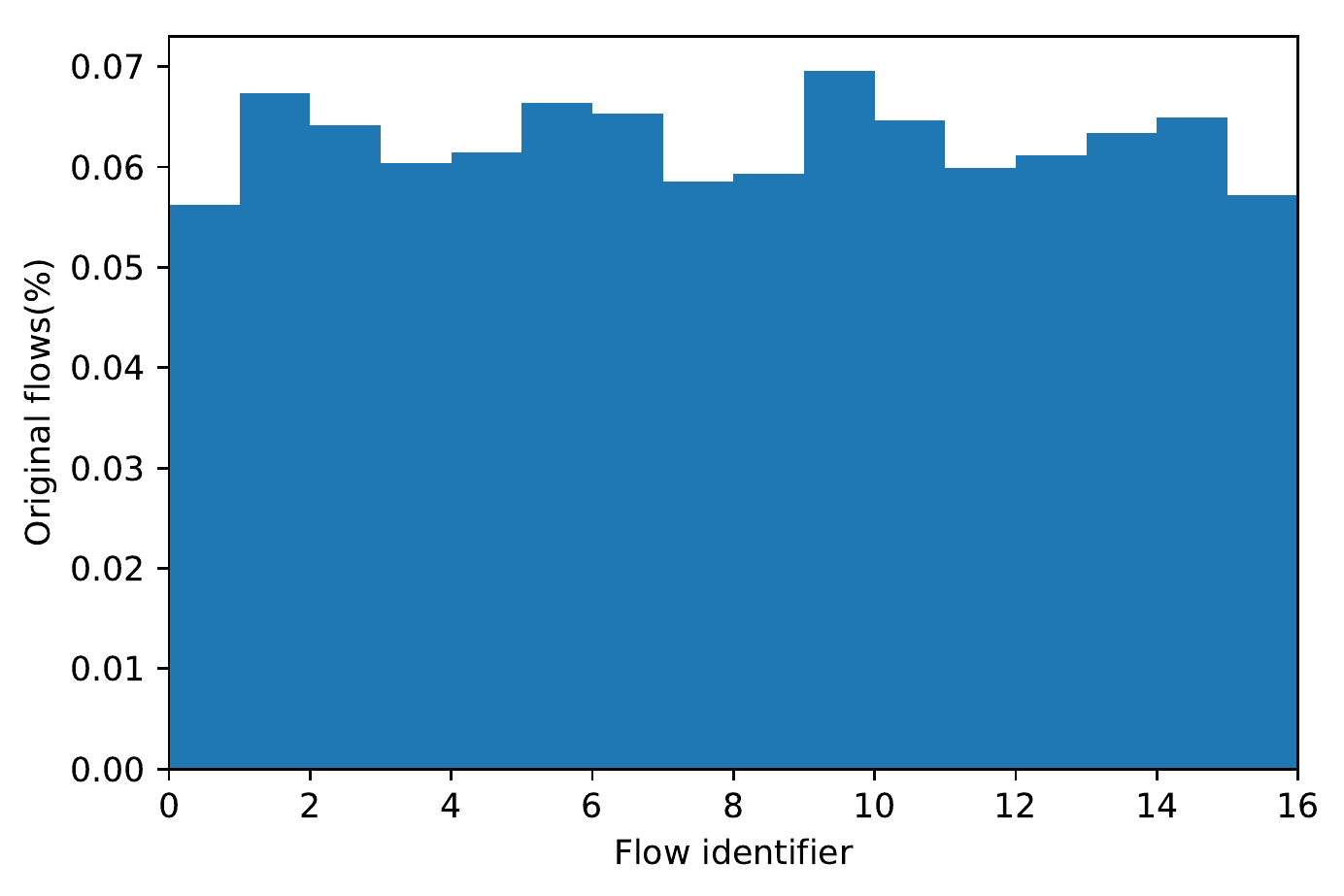}}\hfill
  \subfloat[Last 6 bits]{\includegraphics[width=.49\textwidth]{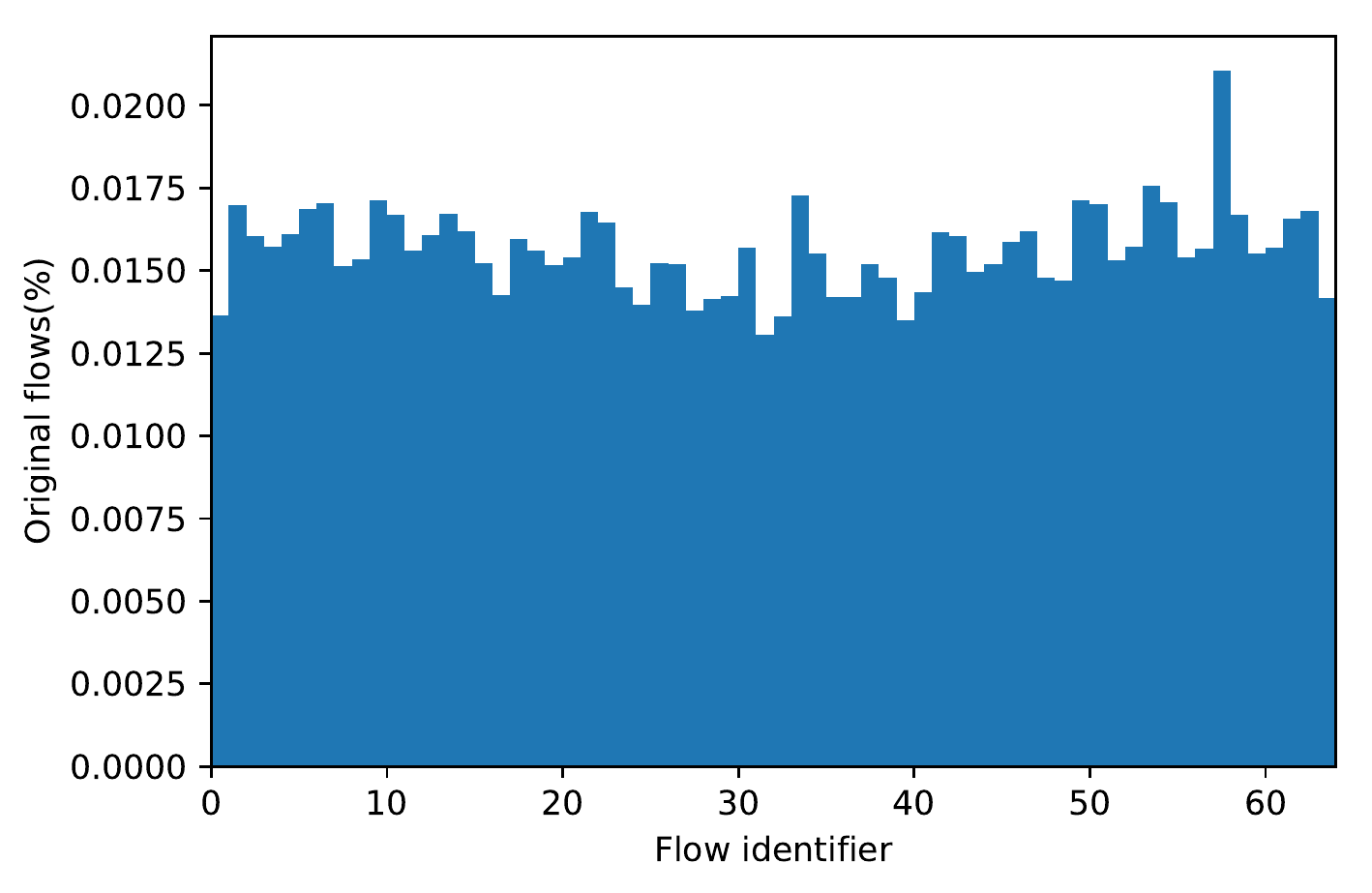}}\\
  \subfloat[Last 10 bits]{\includegraphics[width=.49\textwidth]{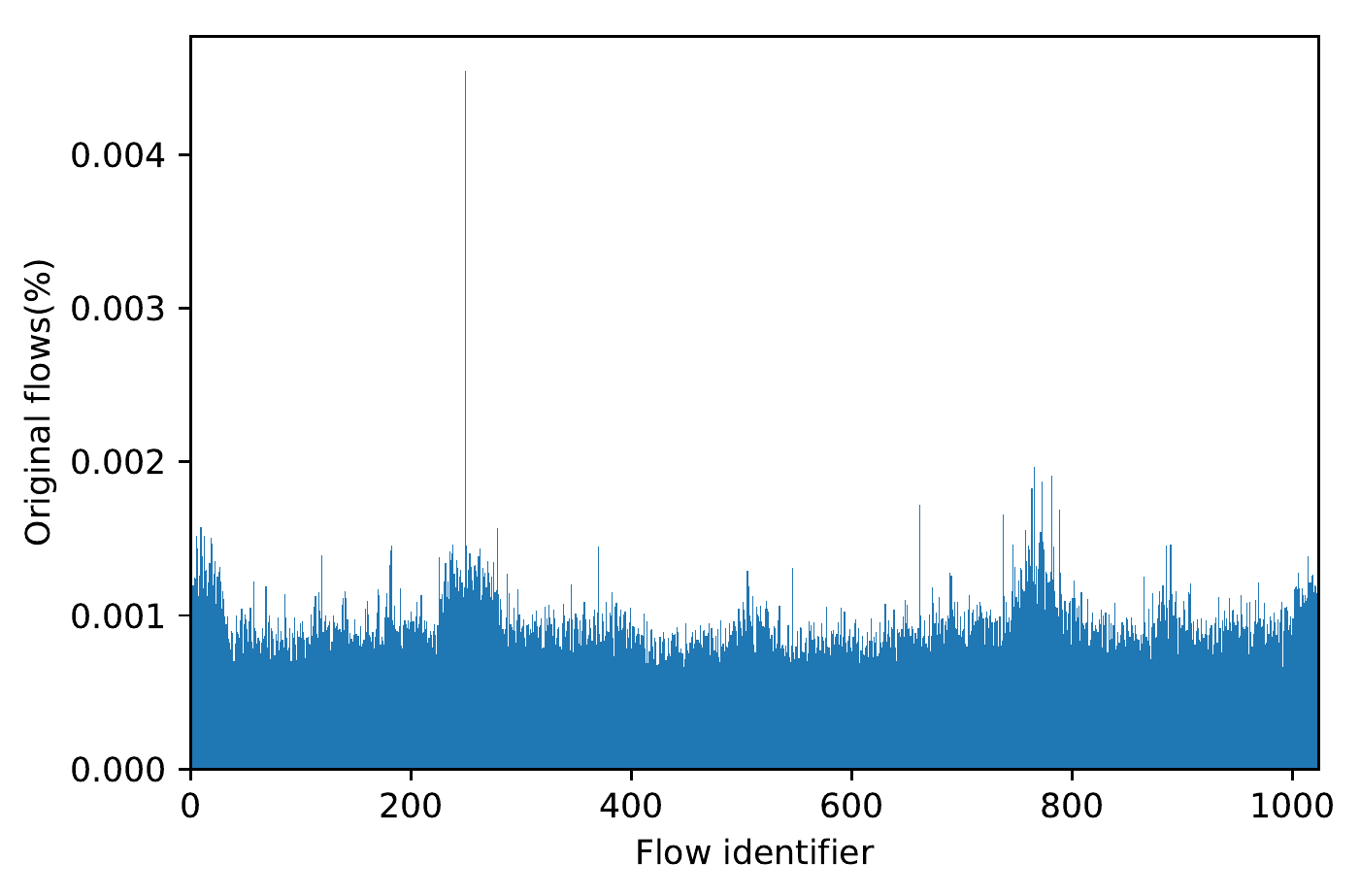}}\hfill
  \subfloat[Last 12 bits]{\includegraphics[width=.49\textwidth]{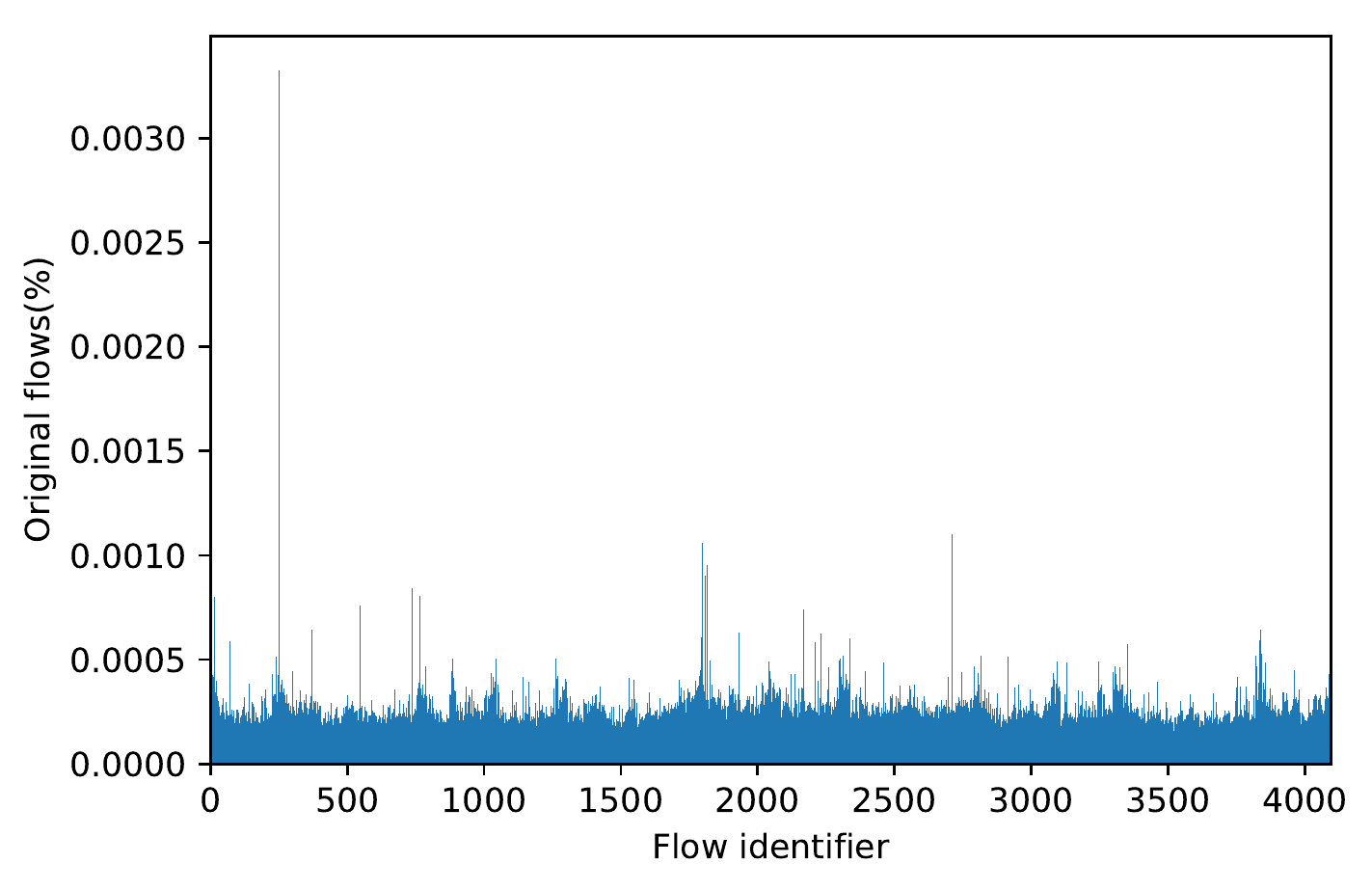}}\\

  \caption{Flow distribution for the last bits of the destination address.}
  \label{fig:hist_dst_comparison}
\end{figure*}
as shown in Fig.~\ref{fig:hist_dst_comparison} and
\begin{table}
  \centering
  \caption{Variance in the number of original flows per aggregated using the
    last bits of the destination IP as flow identifier.}
  \label{tab:var_hist_dst_comparison}
  
  \begin{tabu}{cr}\toprule
    \multicolumn{1}{c}{\textbf{Number of bits}} & \multicolumn{1}{c}{\textbf{Variance}} \\ \midrule
    4 & 7985223 \\
    6 & 954055 \\ 
    8 & 210447 \\ 
    10 & 26446 \\ 
    12 & 4207 \\ \bottomrule
  \end{tabu}
\end{table}
Tab.~\ref{tab:var_hist_dst_comparison}, it also implies a too large number of
aggregated flows being handled by the switch. Consequently, we will use
just the 8 first bits of the destination IP address to define subflows inside
the packets destined to a given MAC address, yielding a maximum of 256~flows.
We see this value as a good trade-off between a small number of flows to be
manageable by the switches and a minimum value of granularity to be able to
spread the traffic among the links.

\subsection{ONOS Application}
\label{sec:onosapp}

Once we have clearly specified how the flows are defined, we proceed to describe the SDN application that we developed. This application has been implemented using ONOS due to it being one of the most supported open source network operating systems. Besides, thanks to the usage of ONOS, our algorithm is agnostic of the SDN forwarding devices and can be directly deployed on any SDN network.

We would like to recall that the application does not work at the packet level
but at the flow level. Consequently, when a packet is received in a switch it
will first look up the installed flow rules and 
it will execute its associated functions whenever a match is found: i.e., 
the specific forwarding port will be selected. Only in the case where a packet
does not yet match any rule among those installed (because it does
not belong to an active flow) it will be sent to the controller. The 
controller will then be responsible for installing the corresponding flow rule
in the switch.
This behavior is the classical \emph{reactive} forwarding application. Indeed, in our application switches are initialized without flow rules and the first packet of each
flow is sent to the controller, which knows how to forward that packet and
instructs the switch to install a flow rule for that flow, so that the
following packets of that flow are forwarded directly by the switch at
line rate, without being sent to the controller.

When a new packet which is to be forwarded by a bundle of EEE links is
received at the controller, the application selects a random port of the
bundle. This is done since the controller lacks any previous information 
about the rate of
this flow.\footnote{Note that if a packet is received at the controller necessarily
this is the first packet of a new flow, hence the controller does not have
previous information about the traffic load of this flow.}

To overcome the limitation of not being able to individually send each packet
to the adequate port as we could do with a water-filling algorithm implemented
at the packet level, our application queries periodically the flow rules
installed in the switches and reorganizes them attempting to minimize the
energy consumption. The flow rules that send traffic to a bundle
are carefully analyzed by our algorithm. First, the application attempts to
estimate the load that each flow will transmit in the next interval. Clearly,
without any other source of information, this prediction must be performed using
the information of the bytes transmitted by this flow in the previous
intervals. This information is provided by the counters that the switch stores
along with each flow rule.\footnote{Actually, SDN devices store counters with the
number of bytes that have matched with each flow.} If our application stores
the number of bytes that each flow has transmitted up to that point, in the
next sampling interval we can calculate the number of bytes transmitted in
that interval as the difference between the total number of bytes at this
point and the previous stored value for that flow. This value of bytes will be
our measure of the traffic of each flow in that interval. A scaling factor is
used for newborn flows (i.e., flows that were not present in the previous
interval) considering the fraction of the interval that each flow has been
active.

Using this information, the algorithm decides which flows will be assigned to
each port, attempting to minimize the overall energy consumption of each
switch, hence minimizing the energy consumption of the whole network. Finally,
this modifications are instructed to the switch, which updates the flow rules
in accordance.

Accordingly, the main tasks of the algorithm are two-fold: first, estimate the
load that each flow will request and secondly, compute an energy-efficient
assignation of the flows to the ports. We will now describe the algorithms
that we have implemented.

\section{Assignment Algorithms}
\label{algorithms}

Once the application has selected the flows, the next task is to decide the
best criteria to assign the identified flows to the set of ports belonging
to the bundle.

In every interval, this algorithm estimates the traffic that each flow will
transmit in the next interval strictly based on the bytes that have been
transmitted by this flow in the previous interval.

\subsection{Greedy Algorithm}

A straightforward way to assign the flows consists on assigning them in order
of decreasing demand, using a new port if the flow does not fit in any of the 
already used flows. The main advantage of such a simple approach is that it draws
few computation resources at the controller. This algorithm is reminiscent of
the classical water-filling approach but the unit of filling is the flow
rather than the packet.

\begin{figure} 
\begin{lstlisting}
allocate_greedy(flows, ports, bound=0) {
  // Hold assigned port for each flow
  flow_allocation[1..$|$flows$|$] = $\varnothing$ 

  /* Sort flows by decreasing load value */
  ordered_flows = sort(load(flows), DECREASING)

  // Initialize occupation of the ports to 0
  port_load[1..$|$ports$|$] = 0
  port_flows[1..$|$ports$|$] = 0

  for flow $\in$ ordered_flows {
    for port $\in$ ports {
      if ((port_flows[port] == 0) ||
          (port_load[port] + load(flows)[flow] 
           $\leq$ 1 - bound/port_flows[port])) {
        // Update port with the load of this flow
        port_load[port] += load(flows)[flow]
        port_flows[port] += 1
        flow_allocation[flow] = port
        break
      }
    }
  }
  
  return flow_allocation
\end{lstlisting}
\caption{Pseudocode for the \emph{Greedy} Algorithms.}
\label{lst:alg2}
\end{figure}
In detail, the flows that forward packets to the bundle are sorted in a decreasing
order based on this estimation of the traffic that will be transmitted. Then, these
flows are sequentially allocated to the ports maximizing the port occupation:
if a flow can be allocated on the port with the highest occupation (i.e., if
the sum of the estimated load of that flow plus the estimated load of the
flows already assigned to that port is less than the capacity of the port) the
flow is assigned to that port; otherwise, the next ports are analogously
evaluated until a port where this flow can be allocated is found. The pseudo-code
for the algorithm is shown in Fig.~\ref{lst:alg2}.

This algorithm is expected to have a good behavior in terms of energy
consumption. Nevertheless, the algorithm does not perform any kind of control
over the amount of packets that need to be queued on each port, requiring very
great buffers (consequently introducing a considerable delay) in order to have
a low percentage of packet losses.

\subsection{Bounded-Greedy Algorithm}

This algorithm is a variation of the previous one with the goal of reducing
the size of the buffer needed on each port for a given packet loss ratio. The
basic operation of the algorithm is the same as in the previous, but instead of
filling the ports to their maximum capacity, we have constrained the
maximum traffic load on any link to a function of the number of flows already
allocated to it.

This way, we allow to reach higher aggregated loads on ports with a high number
of allocated flows, as the aggregated variance of their demand is, in general,
lower than that of ports with only a few flows. The pseudo-code is also
shown in Fig.~\ref{lst:alg2}, where \emph{bound} is the maximum amount of reserved
space that must be left in a port with just one flow allocated to it.

\subsection{Conservative Algorithm}

Unfortunately, the previous algorithms cannot obtain acceptable results in
terms of packet loss for a given buffer size. We have implemented another
energy-efficient algorithm which, as an added benefit, minimizes the length
of the transmission queues.

\begin{figure}
  \centering
  \includegraphics[width=\columnwidth]{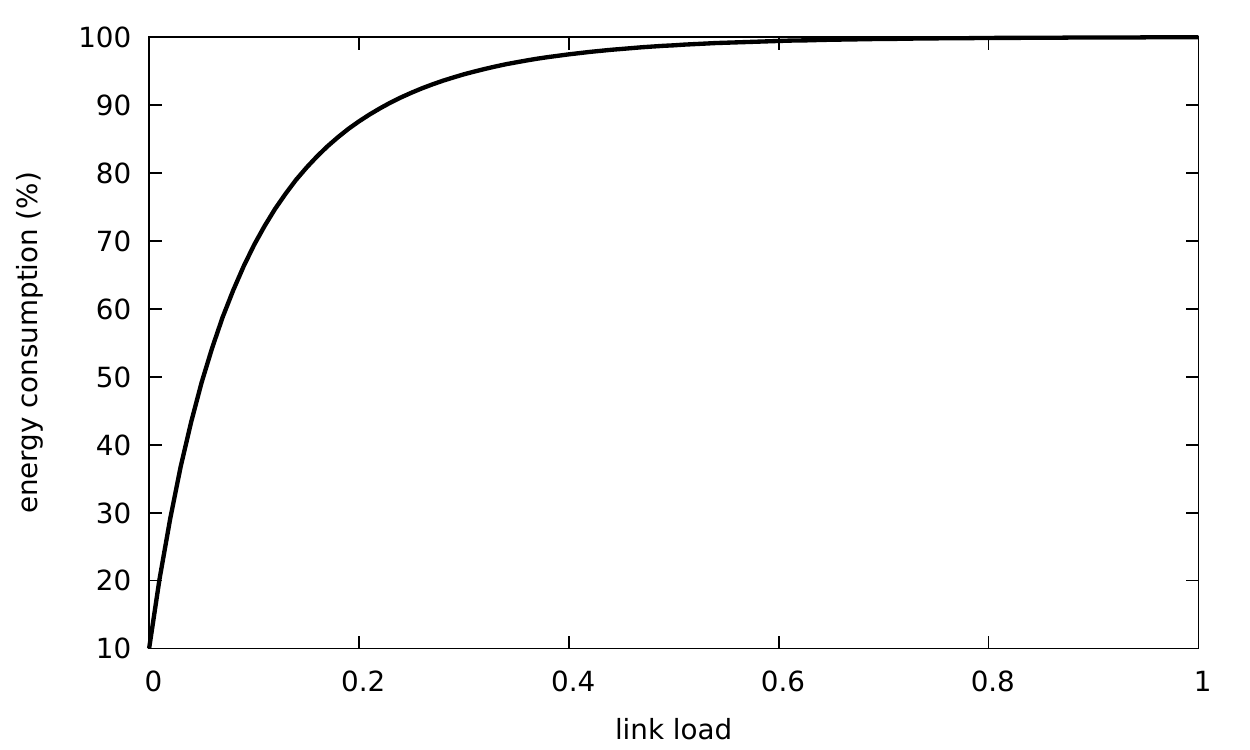}
  \caption{Individual consumption of a \SI{10}{\giga\bit\per\second} IEEE
    802.3az interface.}
  \label{fig:eee_theoretical}
\end{figure}

This algorithm, computes the total estimated traffic load that will
be transmitted through the bundle in the next interval. This value 
puts a lower bound on the number of active ports for the next interval.
Then, flows are spread evenly among all the active links. This clearly minimizes
individual link occupation, but does not follow the water-filling algorithm.
However, this does not matter much, as will be shown later in the results section.
For a port governed by the \emph{frame transmission} algorithm, the energy usage
raises very rapidly with traffic load, see Fig.~\ref{fig:eee_theoretical}.
So, it does not matter that much the load transmitted by each link once it
is higher than about \SI{20}{\percent} of its nominal capacity.

To further avoid packet losses, we do not directly use the estimated load to
calculate the number of used ports, but we first add a safety margin load~\SI{20}{\percent}
to avoid cases where all ports would be used too close to their nominal capacity.

Once we have calculated the number of ports of the bundle that we will be
using, we proceed to minimize the occupation of each port in order to obtain
an homogeneous occupation of all the used ports: not only similar rate but
also similar number of flows. This is easily achieved sorting the flows in a
decreasing order based on the rate estimation and then sequentially assigning
each flow to the port with the lowest occupation among the ones that will be
used for this interval.
\begin{figure} 
\begin{lstlisting}
safety_margin = 20%

allocate_conservative(flows, ports) {
  // Hold assigned port for each flow
  flow_allocation[1..$|$flows$|$] = $\varnothing$ 

  expected_load = sum(load(flows)) + safety_margin
  minimum_ports = ceil(expected_load)

  // Only use the minimum number of ports
  used_ports = ports[1..minimum_ports]

  /* Sort flows by decreasing load value */
  ordered_flows = sort(load(flows), DECREASING)

  // Initialize occupation of the ports to 0
  port_occupation[1..$|$used_ports$|$] = 0

  for flow $\in$ ordered_flows {
    port = get_port_min_occupation(port_occupation)
    // Update port with the load of this flow
    port_occupation[port] += load(flows)[flow]
    flow_allocation[flow] = port
  }

  return flow_allocation

\end{lstlisting}
\caption{Pseudocode for the \emph{Conservative} Algorithm.}
\label{lst:alg3}
\end{figure}
The algorithm is shown in detail in Fig.~\ref{lst:alg3}.

\section{Experimental Results}
\label{results}

\begin{figure}
  \centering
  \includegraphics[width=\columnwidth]{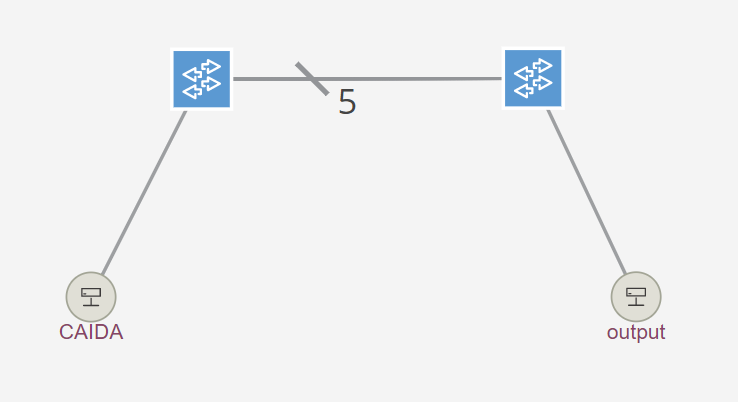}
  \caption{ONOS Web GUI with the setup of the experiment.}
  \label{fig:scenario_topology}
\end{figure}
The validation of the previous algorithms has been carried out on a
scenario composed of two switches interconnected by a 5-link bundle of
10\,GBASE-T interfaces. Fig.~\ref{fig:scenario_topology} shows a snapshot of
the ONOS Web GUI with the analyzed scenario, which has been deployed with
Mininet~\cite{mininet}.

We have employed real traffic traces retrieved from the publicly available
passive monitoring CAIDA dataset~\cite{caida13}, feeding the data to the first
switch so that it had to traverse the link bundle as shown in
Fig.~\ref{fig:scenario_topology}. The trace we have chosen has an average
demand of about \SI{3}{\giga\bit\per\second}, which is relatively low for our
bundle of \SI{50}{\giga\bit\per\second},\footnote{The available traces
  provided by CAIDA are captured on a \SI{10}{\giga\bit\per\second}
  interface.} so we increased the rate tenfold to about
\SI{30}{\giga\bit\per\second} by reducing the inter-arrival times by a
constant factor of ten.

To obtain the energy consumption results, we have proceeded in two
complementary ways. The first one consisted on analytically calculating the
\emph{expected} consumption as the average expected consumption of
its constituting ports. Then, the individual consumption of each port is
calculated as the time average of its instantaneous consumption. As our
algorithms already divide the time in constant intervals, we average over the
consumption in each interval. Finally, the energy consumption on each interval
can be calculated with several well tested models already known in the
literature~\cite{g1g1_eee,marsan11}. In particular, we have
employed~\eqref{eq:energyconsumption}, from the model presented
in~\cite{g1g1_eee}.

\begin{equation}\label{eq:energyconsumption}
  \sigma(\rho_i) = 1 - (1-\soff) (1-\rho_i)
  \frac{%
    \esp{\Toff(\rho_i)}
  }{%
    \esp{\Toff(\rho_i)} + \TS + \TW
  },
\end{equation}
where $\sigma(\cdot)$ is the normalized energy usage, $\rho_i$ is the normalized traffic load on link $i$. We have set
$\soff = 0.1$ according to several estimates provided by different
manufacturers, and $\TS = \SI{2.28}{\us}$ and $\TW = \SI{4.48}{\us}$ as per
the standard~\cite{802.3az}. Besides, assuming frame transmission mode is used
in the IEEE 802.3az interfaces, for Poisson arrivals, we have
\begin{equation}
  \esp{\Toff(\rho)} = \frac{e^{-\mu \rho \TS}}{\mu \rho} 
\end{equation}
where $\mu^{-1}$ is the average packet transmission duration.

We have further verified the results with a IEEE 802.3az simulator, available
for download at~\cite{HystEEE}. To this end we have fed the exact same traffic
that we sent via each port to five instances of the simulator, to then average
the results, obtaining the global consumption. This later result is the one
used in the following figures, as it does not depend on the veracity of the
mathematical models. In any case, the differences were very minor, further
confirming the validity of the models.

Using the above formulas, the theoretical lower bound for the energy
consumption is \SI{78.5}{\percent} when considering a packet size of
\SI{1500}{bytes}. This is achieved when the traffic in the trace, which has an
average rate of \SI{32.5}{\giga\bit\per\second}, is split in the bundle as
follows: 3 ports with \SI{10}{\giga\bit\per\second}, one with
\SI{2.5}{\giga\bit\per\second} and the remaining one completely idle, which
yield 3 ports consuming the \SI{100}{\percent}, one consuming
\SI{83.25}{\percent} and another consuming \SI{10}{\percent}, respectively. We
obtain the global consumption of \SI{78.5}{\percent} as the average of these
five values.

\subsection{Experimental Setup}
Although there exist several simulators such as ns-2 network simulator, we
have decided to implement our custom network simulator in Java, available for
download at \cite{sdn-bundle-simulator}, so that we can share most of the
relevant code
with the ONOS application. 

We are interested mainly in two performance metrics.
Firstly, the overall normalized energy consumption is the main metric
that we have used to validate our algorithms. On the other hand, we have also
measured the packet losses induced by our algorithm for a given buffer size
(in number of packets), i.e., when a buffer in a port is full of packets, new
packets forwarded to that port are discarded.
Moreover, since our algorithm takes effect after the first interval (during the first interval flows are allocated randomly since we do
not have any \emph{a priori} information about the flows) we have decided not
to consider the values of consumption and packet losses of this first
interval, which can be considered as a brief transient state.

To use as a baseline of performance to compare the results of our algorithms with, we have implemented an \emph{equitable} algorithm, 
which just homogeneously spreads the traffic among all the ports in the bundle.

\begin{figure}
  \centering
  \includegraphics[width=\columnwidth]{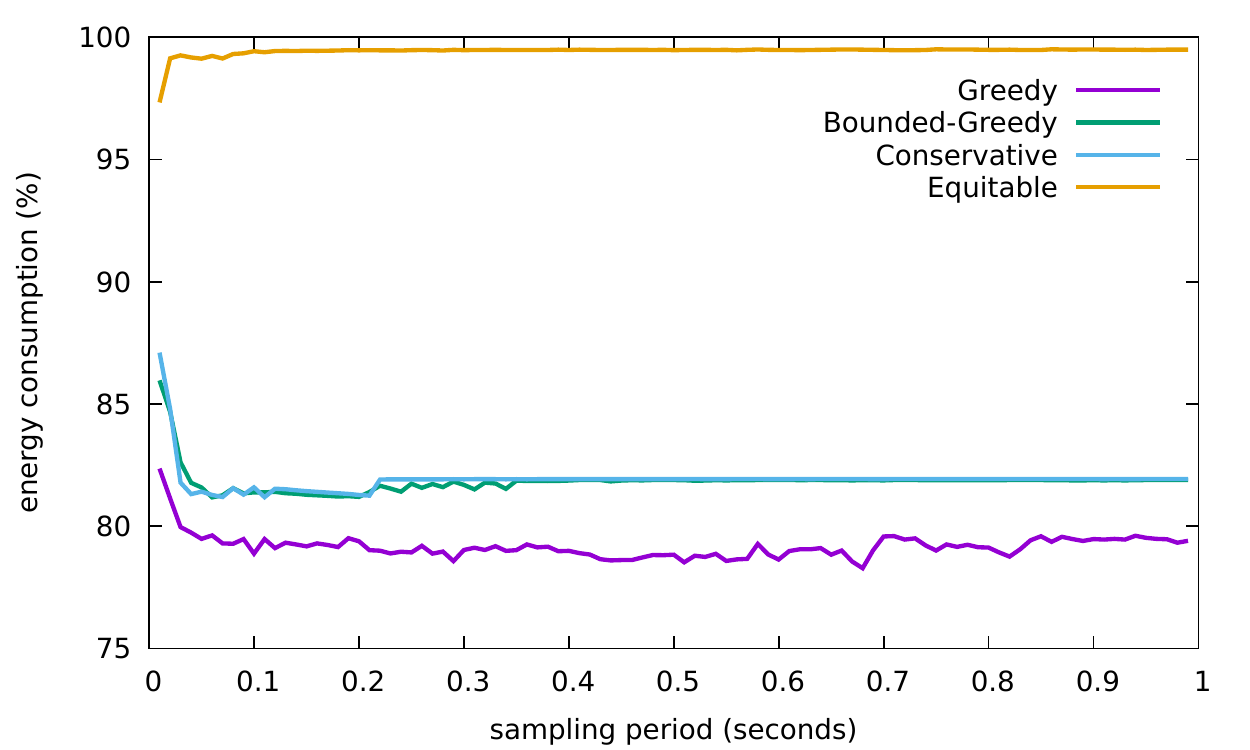}
  \caption{Energy consumption variation with the duration of the sampling period.}
  \label{fig:energy_delay}
\end{figure}
Our first experiment evaluates how the energy consumption of our proposed
algorithms varies with the duration of sampling period.
Fig.~\ref{fig:energy_delay} shows the results of that experiment for a buffer
size of \SI{10000}{packets}.

As expected, the three algorithms outperform the equitable algorithm in terms of energy savings, consuming almost \SI{20}{\percent} less. Besides, we can appreciate that the algorithms results are very close to the analytical minimum bound for the energy consumption. We can also appreciate that the energy savings obtained by the greedy algorithm are slightly higher than those of the other two algorithms, which consume almost the same. In addition, we can notice from Fig.~\ref{fig:energy_delay} that values lower than \SI{0.05}{seconds} yield noticeably worst results than values greater than \SI{0.05}{seconds} in terms of energy consumption.

\begin{figure}
  \centering
  \includegraphics[width=\columnwidth]{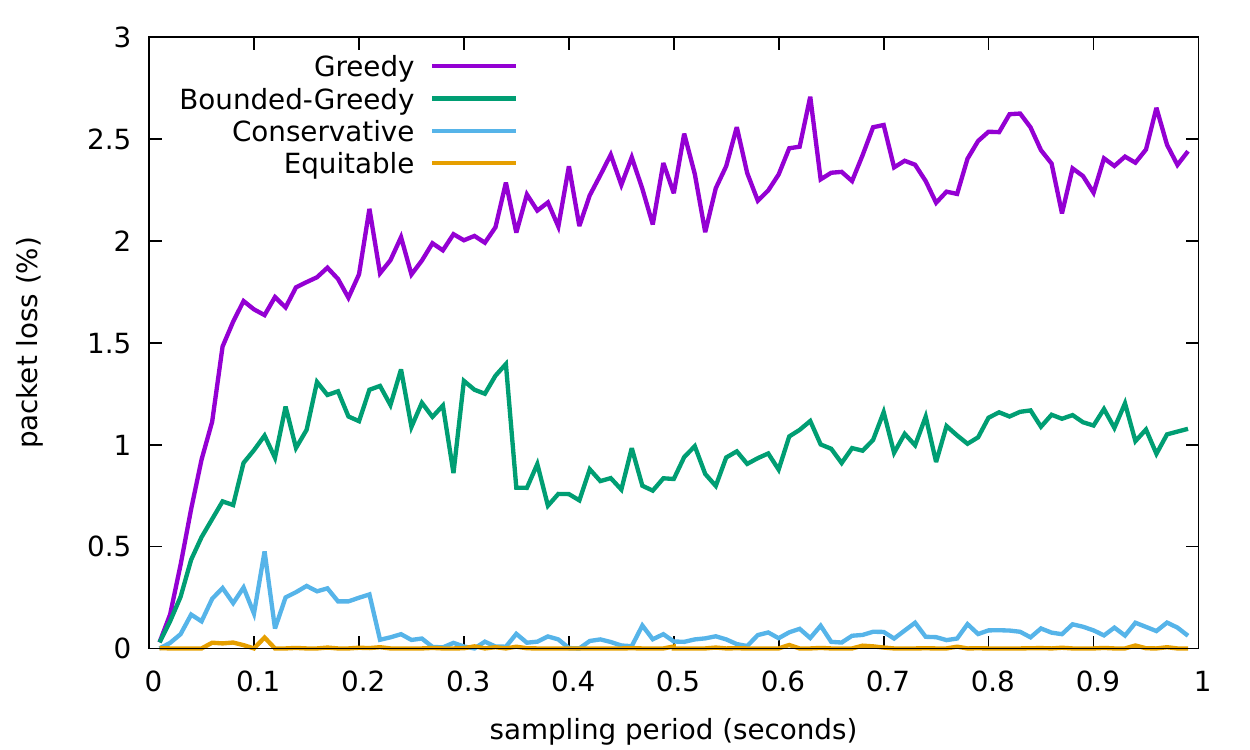}
  \caption{Packet loss percentage variation with the duration of the sampling period.}
  \label{fig:loss_delay}
\end{figure}

\begin{figure}
  \centering
  \includegraphics[width=\columnwidth]{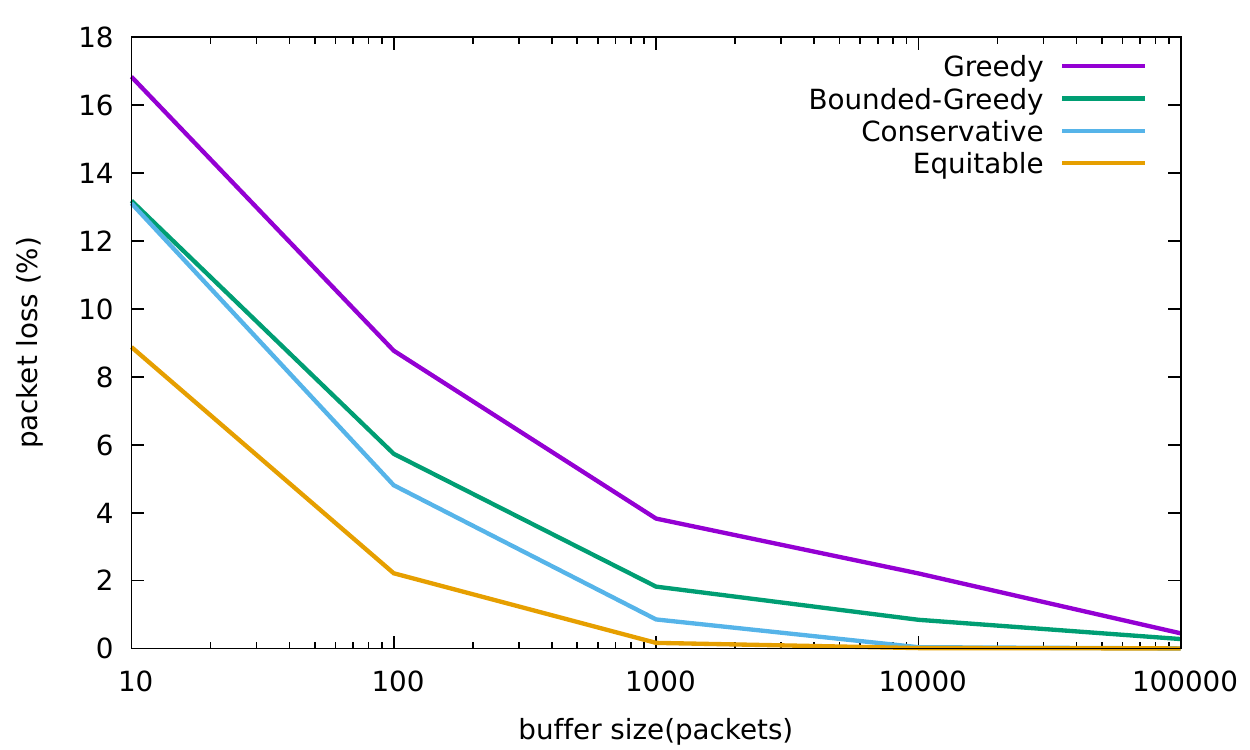}
  \caption{Packet loss percentage variation with the buffer size.}
  \label{fig:loss_buffer}
\end{figure}

The next experiments evaluate the impact on the packet losses induced by our
algorithm. Fig.~\ref{fig:loss_delay} shows the packet loss percentage
variation with the sampling period for a given buffer size of
\SI{10000}{packets}. Fig.~\ref{fig:loss_buffer} represents the packet losses
for different buffer sizes, for a given sampling period of \SI{0.5}{seconds}.

The impact in packet loss depicted in the figures shows how the conservative algorithm outperforms the other two, but without reaching the level of the equitable one. In Fig.~\ref{fig:loss_delay} we can also appreciate that this values exhibited by the conservative algorithm are indeed very close to 0 for almost any value of the sampling period (for a buffer of \SI{10000}{packets}).

Carefully analyzing this graphs we can appreciate that lowering the energy
consumption implies incrementing the packet losses. Therefore, both metrics
cannot be simultaneously optimized. Nevertheless, the improvement in the
energy savings achieved by greedy algorithm with respect to the conservative
one seems not to be worthy, due to the great impact in the packet
loss rate. However, although in view of Fig.~\ref{fig:loss_buffer}, low
values of the sampling period such as \SI{0.01}{seconds} report good values of
packet losses, this is hardly implementable in practice: such a low sampling
period would imply not only sampling the flow tables of every switch 100 times
per second, but also sending up to 256 flow modifications every
\SI{10}{\milli\second}. This is clearly a considerable amount of control
traffic and imposes a huge number of flow modifications per second that would
barely be manageable by the switches. Therefore, for a practical solution
with a negligible performance degradation, we should employ a sampling period of
at least \SI{0.5}{seconds}.
Finally, 
the energy consumption of the bounded-greedy algorithm is almost the same as
the conservative, but the latter is clearly better in terms of packet loss
percentage. Hence, there is no advantage in using the bounded-greedy rather
than the conservative in any way, since the computational complexity of the
three algorithms is equivalent.

To sum up, our energy savings do not come completely free: energy
savings increase traffic delay. Consequently, if the delay must be
bounded to a low value, a low buffer size has to be used, leading to appreciable
packet losses due to the traffic rate variability of the transmitted flows.
The conservative algorithm is able to obtain the minimum traffic delay and packet
losses while obtaining almost identical energy savings. For instance, for 
the analyzed \SI{30}{\giga\bit\per\second} traffic trace, the delay averages
 \SI{270}{\micro\second} and the energy consumption is
\SI{82}{\percent} when using a flow sampling period of \SI{0.5}{\second} and a buffer size of \SI{10000}{packets}. 
For the conservative algorithms, 
this trade-off between delay and energy savings can be tuned via the \emph{safety margin} parameter (which we have set to
\SI{20}{\percent} up to this point). For instance, this same scenario using a
safety margin of \SI{70}{\percent} reduces the average delay to
\SI{150}{\micro\second}, although it elevates the consumption to nearly
\SI{91}{\percent} which is still a \SI{9}{\percent} improvement in the energy
consumption.

\section{Conclusions}
\label{conclusions}

This paper demonstrates the implementation of an energy saving algorithm using
the facilities provided by SDN equipment. We have used the ONOS network
operating system to reduce the energy consumption of an Ethernet link aggregate
between to switches with IEEE~802.3az ports.

The obtained results match those predicted by the packet level model of the
energy saving algorithm we have employed as the basis for our implementation.
Thanks to the usage of ONOS, our algorithm is ready to be deployed in any SDN
network, irrespective of the underlying SDN technology of the equipment
manufactures.

We plan to extend the implementation to cover the case where several link
aggregates are present in the SDN network, to harness the flow
selection work already carried out by the switches for aggregates downstream
to the flows, leveraging ONOS's centralized view of the topology.

\section*{Acknowledgements}

This work was supported by the ``Ministerio de Economía, Industria y
Competitividad'' through the project TEC2017-85587-R of the ``Programa Estatal
de Investigación, Desarrollo e Innovación Orientada a los Retos de la
Sociedad,'' (partly financed with FEDER funds).


\bibliographystyle{IEEEtran}
\bibliography{IEEEfull,sdnbundle.bib}

\begin{thebibliography}{10}
\providecommand{\url}[1]{#1}
\csname url@samestyle\endcsname
\providecommand{\newblock}{\relax}
\providecommand{\bibinfo}[2]{#2}
\providecommand{\BIBentrySTDinterwordspacing}{\spaceskip=0pt\relax}
\providecommand{\BIBentryALTinterwordstretchfactor}{4}
\providecommand{\BIBentryALTinterwordspacing}{\spaceskip=\fontdimen2\font plus
\BIBentryALTinterwordstretchfactor\fontdimen3\font minus
  \fontdimen4\font\relax}
\providecommand{\BIBforeignlanguage}[2]{{%
\expandafter\ifx\csname l@#1\endcsname\relax
\typeout{** WARNING: IEEEtran.bst: No hyphenation pattern has been}%
\typeout{** loaded for the language `#1'. Using the pattern for}%
\typeout{** the default language instead.}%
\else
\language=\csname l@#1\endcsname
\fi
#2}}
\providecommand{\BIBdecl}{\relax}
\BIBdecl

\bibitem{Chiaraviglio2012}
L.~Chiaraviglio, M.~Mellia, and F.~Neri, ``{Minimizing ISP Network Energy Cost:
  Formulation and Solutions},'' \emph{{IEEE/ACM} Transactions on Networking},
  vol.~20, no.~2, pp. 463--476, Apr. 2012.

\bibitem{Jung2014}
D.~Jung, R.~Kim, and H.~Lim, ``{Power-saving strategy for balancing energy and
  delay performance in WLANs},'' \emph{Computer Communications}, vol.~50, pp.
  3--9, Sep. 2014.

\bibitem{Kim2012}
Y.-M. Kim, E.-J. Lee, H.-S. Park, J.-K. Choi, and H.-S. Park, ``{Ant colony
  based self-adaptive energy saving routing for energy efficient Internet},''
  \emph{Computer Networks}, vol.~56, no.~10, pp. 2343--2354, Jul. 2012.

\bibitem{RodriguezPerezb}
M.~Rodríguez~Pérez, M.~Fernández~Veiga, S.~Herrería~Alonso, M.~Hmila, and
  C.~López~García, ``{Optimum Traffic Allocation in Bundled Energy-Efficient
  Ethernet Links},'' \emph{{IEEE} Systems Journal}, p. in press, 2015.

\bibitem{onos}
\BIBentryALTinterwordspacing
``{ONOS - A new carrier-grade SDN network operating system designed for high
  availability, performance, scale-out}.'' [Online]. Available:
  \url{https://onosproject.org/}
\BIBentrySTDinterwordspacing

\bibitem{ontheroad}
K.~Christensen, P.~Reviriego, B.~Nordman, M.~Bennett, M.~Mostowfi, and J.~A.
  Maestro, ``{IEEE 802.3az: the road to Energy Efficient Ethernet},''
  \emph{IEEE Communications Magazine}, vol.~48, no.~11, pp. 50--56, Nov. 2010.

\bibitem{tuysuz2017survey}
M.~F. Tuysuz, Z.~K. Ankarali, and D.~Gözüpek, ``A survey on energy efficiency
  in software defined networks,'' \emph{Computer Networks}, vol. 113, pp.
  188--204, Feb. 2017.

\bibitem{rodrigues2015greensdn}
B.~B. Rodrigues, A.~C. Riekstin, G.~C. Januário, V.~T. Nascimento, T.~C. M.~B.
  Carvalho, and C.~Meirosu, ``{GreenSDN: Bringing energy efficiency to an SDN
  emulation environment},'' in \emph{2015 IFIP/IEEE Int. Symp. Integr. Netw.
  Manag.}\hskip 1em plus 0.5em minus 0.4em\relax IEEE, May 2015, pp. 948--953.

\bibitem{heller2010elastictree}
B.~Heller, S.~Seetharaman, P.~Mahadevan, Y.~Yiakoumis, P.~Sharma, S.~Banerjee,
  and N.~McKeown, ``{Elastictree: Saving energy in data center networks.}'' in
  \emph{Nsdi}, vol.~10, 2010, pp. 249--264.

\bibitem{huong2011ecodane}
T.~Huong, D.~Schlosser, P.~Nam, M.~Jarschel, N.~Thanh, and R.~Pries,
  ``Ecodane—reducing energy consumption in data center networks based on
  traffic engineering,'' in \emph{11th Würzburg Workshop on IP: Joint ITG and
  Euro-NF Workshop Visions of Future Generation Networks (EuroView2011)}, 2011.

\bibitem{mckeown2008openflow}
N.~McKeown, T.~Anderson, H.~Balakrishnan, G.~Parulkar, L.~Peterson, J.~Rexford,
  S.~Shenker, and J.~Turner, ``{OpenFlow: enabling innovation in campus
  networks},'' \emph{ACM SIGCOMM Computer Communication Review}, vol.~38,
  no.~2, pp. 69--74, 2008.

\bibitem{caida13}
\BIBentryALTinterwordspacing
``{The CAIDA UCSD Anonymized 2013 Internet Traces} --- {2013/08/15 13:14:00
  UTC}.'' [Online]. Available:
  \url{https://www.caida.org/data/passive/passive_2013_dataset.xml}
\BIBentrySTDinterwordspacing

\bibitem{mininet}
\BIBentryALTinterwordspacing
``{Mininet: An Instant Virtual Network on your Laptop (or other PC)}.''
  [Online]. Available: \url{http://mininet.org/}
\BIBentrySTDinterwordspacing

\bibitem{g1g1_eee}
S.~Herrería~Alonso, M.~Rodríguez~Pérez, M.~Fernández~Veiga, and
  C.~López~García, ``{A GI/G/1 Model for 10 Gb/s Energy Efficient Ethernet
  Links},'' \emph{{IEEE} Transactions on Communications}, vol.~60, no.~11, pp.
  3386--3395, Nov. 2012.

\bibitem{marsan11}
M.~A. Marsan, A.~F. Anta, V.~Mancuso, B.~Rengarajan, P.~R. Vasallo, and
  G.~Rizzo, ``{A Simple Analytical Model for Energy Efficient Ethernet},''
  \emph{{IEEE} Communications Letters}, vol.~15, no.~7, pp. 773--775, Jul.
  2011.

\bibitem{802.3az}
``{IEEE S}tandard for {I}nformation technology-- {L}ocal and metropolitan area
  networks-- {S}pecific requirements-- part 3: {CSMA/CD A}ccess {M}ethod and
  {P}hysical {L}ayer {S}pecifications {A}mendment 5: {Media Access Control
  Parameters, Physical Layers, and Management Parameters for Energy-Efficient
  Ethernet},'' \emph{{IEEE Std 802.3az-2010 (Amendment to IEEE Std
  802.3-2008)}}, pp. 1--302, Oct. 2010.

\bibitem{HystEEE}
\BIBentryALTinterwordspacing
M.~Rodríguez~Pérez, ``{A Rustified Simulator for 10\,{Gb/s EEE} with
  Configurable Hysteresis}.'' [Online]. Available:
  \url{https://migrax.github.io/HystEEE/}
\BIBentrySTDinterwordspacing

\bibitem{sdn-bundle-simulator}
\BIBentryALTinterwordspacing
P.~Fondo-Ferreiro, ``{SDN Bundle Network Simulator}.'' [Online]. Available:
  \url{https://pfondo.github.io/sdn-bundle-simulator/}
\BIBentrySTDinterwordspacing

\end{thebibliography}

\end{document}